\documentclass[prd,showpacs,showkeys,twocolumn,floatfix]{revtex4} 
\usepackage[]{epsfig,dcolumn}
\usepackage{graphicx}
\usepackage{bm}

\def\IE{{\it i.e.,}}
\def\EG{{\it e.g.}}

\def\l{{\bf l}}

\def\lsim{\mathrel{\rlap{\lower4pt\hbox{\hskip1pt$\sim$}}
   \raise1pt\hbox{$<$}}}
\def\gsim{\mathrel{\rlap{\lower4pt\hbox{\hskip1pt$\sim$}}
    \raise1pt\hbox{$>$}}}

\begin{document}


\title{ Towards a Relativistic Description of Exotic Meson Decays } 

\author{ Nikodem J. Poplawski, Adam P. Szczepaniak, J.T.~Londergan }
\affiliation{ Physics Department and Nuclear Theory Center \\
Indiana University, Bloomington, Indiana 47405 }

\date{\today}

\begin{abstract}

This work analyses hadronic decays of exotic mesons, with a focus 
on the lightest one, the $J^{PC}=1^{-+}$ $\pi_{1}$,
in a fully relativistic formalism, and makes comparisons with 
non-relativistic results. We also discuss Coulomb gauge decays of 
normal mesons that proceed through their hybrid components.
The relativistic spin wave functions of mesons and hybrids are 
constructed based on unitary representations of the Lorentz group. 
The radial wave functions are obtained from phenomenological 
considerations of the mass operator. Fully relativistic results (with 
Wigner rotations) differ significantly from non-relativistic ones.
We also find that the decay channels 
$\pi_{1}\rightarrow\pi b_{1},\,\pi f_{1},\,KK_{1}$ are favored, 
in agreement with results obtained using other models.

\end{abstract}


\maketitle

\section{Introduction}

Strong hadronic decays have been a subject of phenomenological 
investigations for several 
years~\cite{decay1,decay2,decay3,decay4,decay5,decay6}. 
The quark model description of strong decays is based on the assumption that 
decays originate from creation of a quark-antiquark ($Q{\bar Q}$) pair 
in the gluonic field of the decaying meson. The produced  $Q{\bar Q}$ pair 
subsequently recombines with the spectator constituents and hadronizes 
into the final state decay products. Such a picture is consistent with 
the experimental observation that most hadronic decays involve a
minimal number of final state particles in the final states; this requires 
a small number of internal transitions at the quark level.  Furthermore, 
the absence of complicated multi-particle hadronic decays is also 
consistent with a minimal analyticity assumption for the scattering 
amplitude. The absence of complicated multi-particle cuts in the 
s-channel provides justification for a simple resonance Regge pole 
parametrization of the t-channel amplitude over a wide s-channel energy 
range.  

The mechanism of $Q{\bar Q}$ pair production is by itself a
complicated phenomenon, which can in principle be studied through 
lattice gauge simulations. In the strong coupling and/or non-relativistic 
limits, pair production was shown to be similar to electron-positron 
production in a strong uniform electric field (the Schwinger 
mechanism)~\cite{3p0-lattice}. In this case the produced $Q{\bar Q}$ 
pair carries the quantum numbers of the vacuum, \IE unit spin coupled to 
unit relative orbital angular momentum.  
A phenomenological hadron decay model based on such a $Q{\bar Q}$ 
production mechanism is referred to as the $^3P_0$ decay model (the 
spectroscopic notation refers to the quantum numbers of the
produced  $Q{\bar Q}$ pair); this model has been extensively used in 
phenomenological studies of meson and baryon decays.

Within a non-relativistic or constituent quark model Born-Oppeinhemer 
approach, it is assumed that formation of gluonic field distributions 
decouples from the dynamics of the slowly moving constituent quarks. 
Consequently the decay and formation of final state hadrons can be 
studied within a non-relativistic quantum mechanical 
framework~\cite{french-book}. 
For light quarks the non-relativistic approximation can be justified from 
the observation that dynamical chiral symmetry breaking leads to 
massive constituent quarks and transverse gluon excitations with a mass 
gap on the order of 1 GeV~\cite{latice1,latice2}. This heavy, effective mass of gluonic excitations 
 results in  weak mixing between the valence and multi-particle Fock sectors and  
suggests the validity of an approach to the decay 
process in which the pair  production interaction is used only once.

This is also consistent with the characteristics of the experimental data 
on decays discussed earlier.  Since fully dynamical lattice simulations of 
the light hadron resonances are not yet available, such a 
phenomenological approach seems to be a reasonable starting point towards 
a description of the strong decays of light hadrons.

In this paper we discuss some the remaining issues pertaining to models 
based on the ideas presented above. The first issue is the 
question of relativistic effects. Even though a simple non-relativistic 
description appears to be quite successful in predicting decay widths of 
resonance with masses as large as 2-3 GeV, the presence of light quarks 
moving with average velocities a substantial fraction of the speed of 
light naturally raises the question of the validity of the non-relativistic 
reduction. In the case of form factors, it has been demonstrated that 
relativistic effects  in the quark model are in general quite 
large~\cite{ff0,ff1,ff2,ff3}. A study of relativistic effects in  
light-cone quantization has also recently been performed~\cite{as-lc-decays}. 
In the absence of a fully consistent dynamical model, one could argue 
that these effects might somehow be mocked up by effective parameters, 
nevertheless if one seeks a more "universal" constituent quark model 
of hadronic properties, it is essential to address the role of relativistic 
corrections in decays.  

Secondly, as discussed above since the $^3P_0$ decay model can be
related to the characteristics of a Wilson loop, in a consistent
description of a decay process one should consider hadrons including flux
tube degrees of freedom and the effects of flux tube 
breaking~\cite{isgurandpaton,isgurandpaton2}. The majority of
models of normal hadron decays do not include such effects. In those models 
the strong decay amplitude is determined by the quark model constituent 
wave function, multiplied by a form factor representing pair production 
that is independent of the quark distribution in the parent hadron. 
This would not be the case for a general string breaking mechanism.

A simple phenomenological picture of hadrons and their decays in terms of 
quantum mechanical wave functions emerges naturally in a fixed gauge 
approach. For example, in the Coulomb gauge the precursor of flux tube 
dynamics, including string breaking, originates from the non-abelian 
Coulomb potential, which also determines the quark wave 
functions~\cite{ases1,ases2,pwas}.  The string couples to the $Q{\bar Q}$ 
pair via transverse gluon emission and absorption in the standard way. 
This coupling carries $^3S_1$ quantum numbers.  It is interesting to 
investigate whether this coupling, combined  with the flux tube dynamics, 
is consistent with the phenomenological $^3P_0$ picture discussed above. 

\begin{figure}
 \includegraphics[width=2.5in]{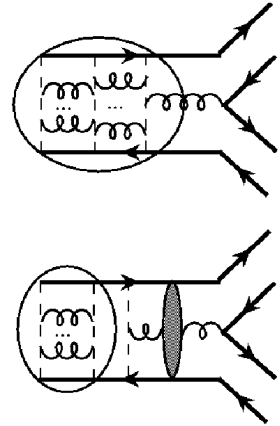}
 \caption{\label{fig1}  Quark flow diagram responsible 
   for strong decays of $Q{\bar Q}g$ hybrids mesons (upper) and $Q{\bar Q}$
   normal mesons (lower). In the upper diagram the dashed lines
   represent the confining non-abelian Coulomb potential. The gluons 
  connecting the Coulomb lines represent formation of the flux tube,
 \EG the gluon string in the ground state. The overall state 
  depicted is enclosed by the solid oval. Emission of a quark-antiquark pair
   from a gluon leads to a re-adjustment of the gluon strings and
   formation of the normal mesons in the final state. 
  In the lower diagram the hybrid meson state appears 
   as an intermediate state in a normal meson decay, which is 
   assumed to proceed via mixing of a $Q{\bar Q}$ pair with  
 virtual excitation of a gluonic string and its subsequent decay. }
 \end{figure}

If one were to attempt a description of decays based on Coulomb
gauge quantization, it is necessary to address the role of the hybrid   
quark-antiquark-gluon configurations, since these appear as 
intermediate states in the decay process as illustrated in 
Fig.~\ref{fig1}.  If such hybrid states also exist as asymptotic states 
they would provide an invaluable insight to the dynamics of confined
gluons. In recent years, evidence has been presented that such states do 
indeed exist, in particular in exotic channels that do not mix with the 
$Q{\bar Q}$ sector~\cite{e852,dzierba,dzierba2}. 
Recently the  non-relativistic $^3P_0$ decay model has been extended
to study hybrid meson decay, modeled via a $Q{\bar Q}$ pair coupled to a 
constituent, non-relativistic string 
(a ``flux-tube'')~\cite{kokosky,closeandpage,esas,pageandas}.   
It is therefore desirable to compare the "universal" decay picture of 
$Q{\bar Q}$ and $Q{\bar Q}g$ mesons emerging from the Coulomb gauge, as 
illustrated in Fig.~\ref{fig1}, with the exotic meson decay phenomenology 
based on models of flux tube breaking.

In this paper we will investigate hybrid meson
decays in a relativistic, Coulomb-gauge motivated description. This is 
a necessary first step towards understanding the phenomenology of
normal meson decays. In the following section we discuss the construction
of the Coulomb wave function for mesons and exotic hybrid mesons. For 
exotic hybrids we concentrate on the $J^{PC}=1^{-+}$ quantum numbers
which are expected for the lightest exotic 
multiplets~\cite{lattice1mp,latex1,latex2,latex3}.  In Section~III we
discuss the role of relativistic effects and give numerical predictions
for various decay modes. In Section~IV we consider decays of normal mesons, 
in particular the $\rho$ and $b_{1}$. In Section~V we summarize our 
results and present future plans.

\section{ Meson and hybrid wave functions } 

 Recently lattice data has become available for static $Q{\bar Q}$ 
 potentials with excited gluonic flux~\cite{latice1,latice2}. In  
Born-Oppeinhemer approximation the lightest  
exotic hybrids, which by definition do not mix with the ground state 
$Q{\bar Q}$ configuration, correspond to states built on top of the 
first excited adiabatic potential. The gluonic configurations can be 
classified according to symmetries of the $Q{\bar Q}$ system, similar to 
the case of a diatomic molecule.  The strong interaction is invariant 
under rotations around the $Q{\bar Q}$ axis, reflection in a plane 
containing the two sources, and with respect to the product of parity
and charge conjugation; thus each configuration can be labeled by the
corresponding eigenvalues denoted by $\Lambda=0,1,{\cdots}$, $Y=\pm 1 $, and 
$PC= \pm 1$ respectively.  In the ground state the gluonic flux tube has
$\Lambda=0$ and in the first excited state it has one unit of spin, 
$\Lambda=1$.  Furthermore in the first excited configuration, lattice 
simulations find $PC=-1$ for the gluon cloud~\cite{latice1,latice2,Swanson:1998kx}

\begin{figure}
 \includegraphics[width=2.5in]{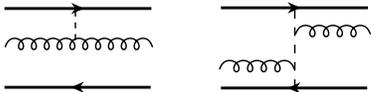}
 \caption{\label{fig2}  Two-body (left) and three-body (right) potential 
between the transverse gluon and the static $Q{\bar Q}$ sources. The dashed 
line represents the expectation value of the non-abelian Coulomb potential.}
 \end{figure}

In the Coulomb gauge the quantum numbers of gluonic states can be 
associated with those of the extra transverse gluon in the presence of 
a static $Q{\bar Q}$ source. This is because the 
transverse gluons are dressed~\cite{ases2,rein}, and on average behave as 
constituents with effective mass 
$m_g \sim 600\mbox{ MeV}$\cite{glue,ases5}. Thus low lying excited 
gluonic states are expected to have a small number of transverse gluons. 
The flux tube itself is expected to emerge from the strong coupling 
of transverse gluons to the Coulomb potential.  The transverse gluon wave 
function can be obtained by diagonalizing the net quark-antiquark-gluon 
interactions shown in Fig.~\ref{fig2} (in addition to the gluon 
kinetic energy). 

 The three-body interaction shown in Fig.~\ref{fig2} is of particular 
relevance.  If the gluon is in a relative $S$-wave with respect to the 
$Q{\bar Q}$ system it has a net $PC=+1$, which is opposite to the lattice 
results for the first excited configuration.  However, the transverse gluon
has a gradient coupling to the Coulomb potential, thus a $P$-wave 
transverse gluon receives no energy shift from this coupling and the 
energy of the $S$-wave gluon state is increased.  In the Coulomb gauge
picture, the shift of the $S$-wave state via this 3-body interaction 
may be the cause of the inversion of $S-P$ levels seen on the 
lattice~\cite{adel}.  A quantitative analysis of this effect will be the 
subject of a separate investigation.  In the following, when considering  
the wave function of an exotic hybrid we will assume that the transverse 
gluon is in a $P$-wave relative to the $Q{\bar Q}$ state.

\subsection{Mesons as $Q\bar{Q}$ bound states}

In this analysis we do not solve the Coulomb gauge QCD Hamiltonian to 
obtain meson wave functions.  Instead, we use the general transformation 
properties under the remaining kinematical symmetries (rotations and 
translations) to construct the states.  Finally we employ Lorentz 
transformations as for non-interacting constituents to obtain 
meson wave functions for finite center of mass {\it (c.o.m.)} momenta that 
are required in decay calculations. This is clearly an approximation which 
cannot be avoided without solving the dynamical equations for the boost 
generators~\cite{ff1,ff2}. In the following section we examine some 
single particle observables to investigate this approximation. 

For a system of non-interacting particles the spin wave function is 
constructed as an element of an irreducible, unitary representation of 
the Poincare group~\cite{gs,fk,bt,bkt,kt}.  We will assume isospin symmetry 
$m_{u,d}=m$ and treat the light flavors first.  The generalization to 
strange and heavier mesons will be given at the end of this section. 

In the rest frame of a quark-antiquark pair the quark momenta are given by
\begin{equation}
\l^{\mu}_{q}=(E(m,{\bf q}),{\bf q}),\,\,l^{\mu}_{\bar{q}}= 
  (E(m,-{\bf q}),-{\bf q}), 
\end{equation}
and the normalized spin-0 and spin-1 wave function corresponding to 
$J^{PC}=0^{-+}$ and $J^{PC}=1^{--}$ are given by the Clebsch-Gordan 
coefficients,
\begin{eqnarray} 
& & \Psi_{q\bar{q}}({\bf q},{\bf l}_{q\bar{q}}=0,\sigma_{q},\sigma_{\bar{q}})=
 \langle \frac{1}{2},\sigma_{q};\frac{1}{2},\sigma_{\bar{q}}|0,0 
  \rangle = \nonumber \\ 
 & & =\frac{[i\sigma_{2}]_{\sigma_{q}
 \sigma_{\bar{q}}}}{\sqrt{2}} 
 \label{0cg}
\end{eqnarray}
and
\begin{eqnarray} 
& & \Psi^{\lambda_{q\bar q}}_{q\bar{q}}({\bf q},
  {\bf l}_{q\bar{q}}=0,\sigma_{q},\sigma_{\bar{q}})=
 \langle \frac{1}{2},\sigma_{q};\frac{1}{2},
  \sigma_{\bar{q}}|1,\lambda_{q\bar q} \rangle = \nonumber \\
 & & =\frac{[\sigma^i i\sigma_{2}]_{\sigma_{q}\sigma_{\bar{q}}}}{\sqrt{2}}\epsilon^{i}(\lambda_{q\bar{q}}), 
 \label{1cg}
\end{eqnarray}
respectively.  These can also be expressed in terms of Dirac spinors as 
\begin{equation}
 \Psi_{q\bar{q}}({\bf q},{\bf l}_{q\bar{q}}=0,\sigma_{q},\sigma_{\bar{q}})
  =\frac{1}{\sqrt{2}m_{q\bar{q}}}\bar{u}({\bf q},\sigma_{q})
  \gamma^{5}v(-{\bf q},\sigma_{\bar{q}}) 
\label{eq:00r}
\end{equation}
and
\begin{eqnarray}
& & \Psi^{\lambda_{q\bar{q}}}_{q\bar{q}}({\bf q},{\bf l}_{q\bar{q}}=0,
  \sigma_{q},\sigma_{\bar{q}})  \nonumber \\
& & =\frac{1}{\sqrt{2}m_{q\bar{q}}}\bar{u}({\bf q},\sigma_{q})
  \Bigl[\gamma^{i}-\frac{2q^{i}}{m_{q\bar{q}}+2m}\Bigr]v(-{\bf q},
  \sigma_{\bar{q}})\epsilon^{i}(\lambda_{q\bar{q}}), \nonumber \\
\label{eq:01r}
\end{eqnarray}
where $m_{q\bar{q}}$ is the invariant mass of the $Q\bar{Q}$ pair,
$m_{q\bar q}  = E(m,{\bf q}) + E(m, -{\bf q}) =  2 \sqrt{m2 + {\bf q}^2}$,  
and $\epsilon^{i}(\lambda_{q\bar{q}})$ are the polarization vectors 
corresponding to spin $1$ quantized along the z-axis.  The boosted spin 
functions for mesons are summarized in Appendix A. 
The wave function of a $Q\bar{Q}$ system moving with a total momentum 
${\bf l}_{q\bar{q}}={\bf l}_{q}+{\bf l}_{\bar{q}}\ne 0$ is given by
\begin{eqnarray}
& & \Psi^{\lambda_{q\bar{q}}}_{q\bar{q}}({\bf q},{\bf l}_{q\bar{q}},
  \lambda_{q},\lambda_{\bar{q}})=
 \sum_{\sigma_{q},\sigma_{\bar{q}}}\Psi^{\lambda_{q\bar{q}}}_{q\bar{q}}
 ({\bf q},{\bf l}_{\bar{q}}=0,\sigma_{q},\sigma_{\bar{q}}) \nonumber \\
& & \times D^{\ast(1/2)}_{\lambda_{q}\sigma_{q}}({\bf q},
 {\bf l}_{q\bar{q}})D^{(1/2)}_{\lambda_{\bar{q}}\sigma_{\bar{q}}}(-{\bf q},
 {\bf l}_{q\bar{q}}),
\label{eq:boost}
\end{eqnarray}
where the Wigner rotation matrix 
$D^{(1/2)}_{\lambda\lambda'}({\bf q},{\bf P})$ corresponds to a boost 
with ${\bf \beta}\gamma={\bf P}/M$.  One can show (see Appendix A) that 
the corresponding wave functions can be written in terms of
covariant amplitudes, 
 \begin{equation}
 \Psi_{q\bar{q}}({\bf q},{\bf l}_{q\bar{q}},\lambda_{q},\lambda_{\bar{q}})
 =\frac{1}{\sqrt{2}m_{q\bar{q}}}\bar{u}({\bf l}_{q},
 \lambda_{q})\gamma^{5}v({\bf l}_{\bar{q}},\lambda_{\bar{q}})
\label{eq:00}
\end{equation}
and 
\begin{eqnarray}
& & \Psi^{\lambda_{q\bar{q}}}_{q\bar{q}}({\bf q},{\bf l}_{q\bar{q}},
 \lambda_{q},\lambda_{\bar{q}})=  \nonumber \\
& & = -\frac{\epsilon_{\mu}({\bf l}_{q\bar{q}},\lambda_{q\bar{q}})}
 {\sqrt{2}m_{q\bar{q}}}\bar{u}({\bf l}_{q},\lambda_{q})
 \Bigl[\gamma^{\mu}-\frac{l^{\mu}_{q}-l^{\mu}_{\bar{q}}}{m_{q\bar{q}}+2m}
 \Bigr]v({\bf l}_{\bar{q}},\lambda_{\bar{q}}), \nonumber \\
\label{eq:01}
\end{eqnarray}
where $\epsilon^{\mu}({\bf l}_{q\bar{q}},\lambda_{q\bar{q}})$ are obtained 
from $(0,\epsilon^{i}(\lambda_{q\bar{q}}))$ through a Lorentz boost with 
${\bf \beta}\gamma={\bf l}_{q\bar{q}}/m_{q\bar{q}}$. 
Obviously ${\bf l}_{q}=\Lambda(0\rightarrow{\bf l}_{q\bar{q}}){\bf q}$ 
and ${\bf l}_{\bar{q}}=\Lambda(0\rightarrow{\bf l}_{q\bar{q}})(-{\bf q})$.

By coupling Eqs.~($\ref{eq:00r}$) or ($\ref{eq:01r}$) for 
${\bf l}_{q\bar{q}}=0$ with one unit of the orbital angular momentum $L=1$ 
and then making the boost in Eq.~($\ref{eq:boost}$), one obtains respectively 
the spin wave function for the quark-antiquark pair with quantum numbers 
$J^{PC}=1^{+-}$,
\begin{equation}
\Psi^{J,\lambda_{q\bar{q}}}_{q\bar{q}}=\frac{1}{\sqrt{2}m_{q\bar{q}}
 ({\bf l}_{q},{\bf l}_{\bar{q}})}\bar{u}({\bf l}_{q},\lambda_{q})
 \gamma^{5}v({\bf l}_{\bar{q}},\lambda_{\bar{q}})Y_{1\lambda_{q\bar{q}}}
 (\bar{{\bf q}}),
\label{eq:10}
\end{equation}
or for $J^{PC}=0^{++}$, $1^{++}$ and $2^{++}$,
\begin{eqnarray}
& & \Psi^{\lambda_{q\bar{q}}}_{q\bar{q}}=-\sum_{\lambda,l}
 \frac{1}{\sqrt{2}m_{q\bar{q}}}\bar{u}({\bf l}_{q},\lambda_{q})
 \Bigl[\gamma^{\mu}-\frac{l^{\mu}_{q}-l^{\mu}_{\bar{q}}}{m_{q\bar{q}}+2m}
 \Bigr] \nonumber \\
& & \times v({\bf l}_{\bar{q}},\lambda_{\bar{q}})\epsilon_{\mu}
 ({\bf l}_{q\bar{q}},\lambda)Y_{1l}(\bar{{\bf q}})
 \langle 1,\lambda;1,l|J,\lambda_{q\bar{q}} \rangle ,
\label{eq:11}
\end{eqnarray}
with ${\bf q}=\Lambda({\bf l}_{q\bar{q}}\rightarrow0){\bf l}_{q}$.
In order to construct meson spin wave functions for higher orbital angular 
momenta $L$ between the $Q{\bar Q}$ pair one need only to replace $Y_{1l}$ 
with $Y_{Ll}$.  For consistency, we should add the constant factor $Y_{00}$ 
to wave functions with $L=0$.

Now we can proceed with the construction of meson states characterized by 
momentum ${\bf P}$, spin $\lambda_{q\bar{q}}$, and isospin $I$. 
The $\pi\,(I=1)$ and $\eta\,(I=0)$ states ($J^{PC}=0^{-+}$) are constructed 
in terms of the annihilation and creation operators:
\begin{widetext}
\begin{eqnarray}
& &  |M({\bf P},I,I_{3})\rangle\,=\sum_{all\,\,\lambda,c,f}\int\frac{d^{3}
 {\bf p}_{q}d^{3}\,{\bf  p}_{\bar{q}}}{(2\pi)^{6}2E(m,{\bf p}_q) 
 2E(m,{\bf p}_{\bar q}) } 
 2(E(m_q,{\bf p}_q)+E(m_{\bar q},{\bf p}_{\bar q} ) )
 \cdot(2\pi)^{3}\delta^{3}({\bf p}_{q}+{\bf p}_{\bar{q}}-{\bf P}) \nonumber \\
& & \times\frac{1}{N(P)}\frac{\delta_{c_{q}c_{\bar{q}}}}{\sqrt{3}}
 \frac{F(I,I_{3})_{f_{q}f_{\bar{q}}}}{\sqrt{2}}\Psi_{q\bar{q}}({\bf p}_{q},
 {\bf p}_{\bar{q}},\lambda_{q},\lambda_{\bar{q}}) 
\psi_{L}(m_{q\bar{q}}({\bf p}_{q},{\bf p}_{\bar{q}})/\mu)\,
 b^{\dag}_{{\bf p}_{q}\lambda_{q}f_{q}c_{q}}d^{\dag}_{{\bf p}_{\bar{q}}
 \lambda_{\bar{q}}f_{\bar{q}}c_{\bar{q}}}|0\rangle .
\label{eq:pieta}
\end{eqnarray}
\end{widetext}
In the above $\Psi_{q\bar{q}}$ is the spin-0 wave function of 
Eq.~($\ref{eq:00}$), written explicitly in terms of the momenta 
${\bf p}_{q}$ and 
${\bf p}_{\bar{q}}$ instead of the relativistic relative momentum 
 ${\bf q}$ and the {\it c.o.m.} momentum ${\bf P}={\bf l}_{q\bar{q}}$, 
given by 
\begin{equation}
{\bf p}_q = {\bf q} + {{ ({\bf q} \cdot {\bf P}){\bf P}} \over 
 {E(m_{q\bar q},{\bf P}) (m_{q\bar q} 
 + E(m_{q\bar q},{\bf P}) ) } } + {{E(m, {\bf q})} \over {m_{q\bar q}}} 
 {\bf P}
\end{equation}
and
\begin{equation}
{\bf p}_{\bar q} = -{\bf q} - {{( {\bf q} \cdot {\bf P}){\bf P}} \over 
 {E(m_{q\bar q},{\bf P}) (m_{q\bar q} 
 + E(m_{q\bar q},{\bf P}) ) } } + {{E(m, {\bf q})} \over {m_{q\bar q}}} 
 {\bf P}.
\end{equation}
In Eq.~(\ref{eq:pieta}), $c$, $f$ and $I_{3}$ denote respectively 
the color, flavor, and third component of isospin.  $\psi_{L}$ represents the 
orbital wave function resulting from the quark-antiquark interaction that 
leads to a bound state (meson).  We assume that this function depends only 
on the invariant mass $m_{q\bar{q}}$.  The normalization constant $N$ 
(with $P=|{\bf P}|$) is fixed by
\begin{equation}
 \langle {\bf P},\alpha|{\bf P}',\alpha' \rangle \,\,=(2\pi)^{3}2E(m_{M},
 {\bf P})\delta^{3}({\bf P}-{\bf P}')\delta_{\alpha\alpha'},
\label{eq:norm}
\end{equation}
where $m_{M}$ is the meson mass and $\alpha$ represents spin and isospin. 
The parameter $\mu$ is a scalar function of the meson quantum numbers.  
Finally, $F(I,I_{3})$ is a $2\times 2$ isospin matrix ($f=1$ for $u$ and 
$f=2$ for $d$):
\begin{equation}
F(0,0)=I,\,\,\,F(1,I_{3})=\sigma^{i}\epsilon^{i}(I_{3}).
\end{equation}
The flavor structure of the $\eta$ state (as well as other isospin zero 
mesons) was chosen as a linear combination 
$\frac{1}{\sqrt{2}}(|u\bar{u}\rangle+\,|d\bar{d}\rangle)$, although in 
general those states are linear combinations 
$\cos(\phi) [ |u\bar{u}\rangle+\,|d\bar{d}\rangle]/\sqrt{2} + \sin(\phi) 
|s\bar{s}\rangle$. 
The $|s\bar{s}\rangle$ does not contribute to the amplitude of the decay 
of the $\pi_{1}$ and therefore may be neglected in calculations, provided 
this amplitude is multiplied by a factor $\cos(\phi)$. 

Similarly the $\rho\,(I=1)$ and $\omega,\,\phi\,(I=0)$ states 
($J^{PC}=1^{--}$) are given by Eq.\ ($\ref{eq:pieta}$), but instead of 
$\Psi_{q\bar q}$ in Eq.\ ($\ref{eq:00}$) one must use 
 $\Psi_{q\bar{q}}^{\lambda_{q\bar{q}}}$ given in Eq.\ ($\ref{eq:01}$). 
The $b_{1}\,(I=1)$ and $h_{1}\,(I=0)$ states ($J^{PC}=1^{+-}$) contain the 
wave function of Eq.\ ($\ref{eq:10}$).  Finally, the $a\,(I=1)$ and 
$f\,(I=0)$ states ($J^{PC}=0,1,2^{++}$) correspond to Eq.\ ($\ref{eq:11}$).

The above results can be straightforwardly generalized to the case where 
$m_{q}$ and $m_{\bar{q}}$ are different, for example to decays into mesons 
with one strange quark ($I=1/2$).  The spin wave function for a 
quark-antiquark pair in a $J^{P}=0^{-}$ state is
\begin{equation}
\Psi_{q\bar{q}}({\bf l}_{q},{\bf l}_{\bar{q}},\lambda_{q},\lambda_{\bar{q}})
 =\frac{1}{\sqrt{2}{\tilde{m}}_{q\bar{q}}}\bar{u}(m_{q},{\bf l}_{q},
 \lambda_{q})\gamma^{5}v(m_{\bar{q}},{\bf l}_{\bar{q}},\lambda_{\bar{q}}),
\label{eq:00s}
\end{equation}
where $\tilde{m}_{q\bar{q}}=\sqrt{m_{q\bar{q}}^{2}-(m_{q}-m_{\bar{q}})^{2}}$, 
and the $K$-meson states are given by Eq.~($\ref{eq:pieta}$), with an 
appropriate definition of the matrix $F$.  In Eq.~($\ref{eq:pieta}$), if 
$\Psi_{q\bar{q}}$ in Eq.~($\ref{eq:00s}$) is replaced by
\begin{eqnarray}
& & \Psi^{\lambda_{q\bar q}}_{q\bar{q}}=-\frac{1}{\sqrt{2}
 {\tilde{m}}_{q\bar{q}}}\bar{u}(m_{q},{\bf l}_{q},\lambda_{q})
 \Bigl[\gamma^{\mu}-\frac{l^{\mu}_{q}-l^{\mu}_{\bar{q}}}
 {m_{q\bar{q}}+m_{q}+m_{\bar{q}}}\Bigr] \nonumber \\
& & \times v(m_{\bar{q}},{\bf l}_{\bar{q}},\lambda_{\bar{q}})
 \epsilon_{\mu}({\bf l}_{q\bar{q}},\lambda_{q\bar{q}}),
\label{eq:01s}
\end{eqnarray} 
then one obtains the $K^{\ast}$-meson states ($J^{P}=1^{-}$). 
The wave functions of strange mesons with non-zero angular momentum (such 
as $J^{P}=1^{+}$) can be obtained by coupling with the spherical harmonics.

So far we have treated mesons as non-interacting $Q\bar{Q}$ pairs. 
The interaction between a quark and an antiquark enters through the 
Hamiltonian $H=P^{0}$ and the boost generators of the Poincare group $M^{0i}$.
It is possible to produce models of interactions for a fixed number of 
constituents that preserve the Poincare algebra following the prescription 
of Bakamjian and Thomas~\cite{bt,gs}.  Unfortunately such a construction 
does not guarantee that physical observables, \EG  current matrix 
elements and decay amplitudes will obey relativistic covariance. 
In any case one deals with phenomenological models of the quark dynamics,  
therefore we follow the common practice of employing a simple (Gaussian) 
parametrization of the orbital wave functions with one scale parameter 
$\mu$ related to the size of the meson, 
\begin{equation}
 \psi_{L}(m_{q\bar{q}}/\mu)=e^{-m^{2}_{q\bar{q}}/8\mu^{2}}. 
 \end{equation}
In the non-relativistic limit (where the Wigner rotation may be ignored) all 
the spin wave functions for regular mesons simply reduce to the 
Clebsch-Gordan coefficients that we started from.

\subsection{Hybrid mesons as $Q\bar{Q}g$ bound states}

As we discussed previously, in the exotic hybrid meson wave function the 
gluon is expected to have one unit of orbital angular momentum with respect 
to the $Q{\bar Q}$ pair.  In the rest frame of the 3-body system, where the 
$Q\bar{Q}$ pair moves with momentum $-{\bf Q}$ and the transverse gluon with 
momentum ${\bf +Q}$, the total spin wave function of the hybrid is obtained 
by coupling the $Q\bar{Q}$ spin-1 wave function of Eq.~($\ref{eq:01}$) and 
the gluon wave function ($J^{PC}=1^{--}$) to total spin $S=0,1,2$ and 
$J^{PC}=0^{++},1^{++},2^{++}$ states respectively.  The exotic meson wave 
function with $J^{PC}=1^{-+}$ is then obtained by adding one unit of  
orbital angular momentum between the gluon and the $Q{\bar Q}$ pair: 
\begin{eqnarray}
& & \Psi^{\lambda_{ex}}_{q\bar{q}g(S)}(\lambda_{q},\lambda_{\bar{q}},
 \lambda_{g})=\sum_{\lambda_{q\bar{q}},\sigma,M,l}
 \Psi^{\lambda_{q\bar{q}}}_{q\bar{q}}({\bf q},-{\bf Q},\lambda_{q},
 \lambda_{\bar{q}}) \nonumber \\ 
& & \times Y_{1l}(\bar{{\bf Q}})\langle 1,\lambda_{q\bar{q}};1,
 \sigma|S,M \rangle D^{(1)\ast}_{\lambda_{g}\sigma}(\bar{{\bf Q}}) 
 \langle S,M;1,l|1,\lambda_{ex} \rangle. \nonumber \\
\label{eq:exo}
\end{eqnarray}
The spin-1 rotation matrix $D^{(1)}$, representing the gluon spin wave 
function, relates the transverse gluon states in the helicity basis 
$\sigma(=\pm1)$ to the basis described by spin $\lambda_{g}$ quantized 
along a fixed z-axis.  The Clebsch-Gordan coefficients and the spherical 
harmonic in Eq.~(\ref{eq:exo}) can be expressed in terms of the 
polarization vectors, and the action of the rotation matrix on the gluon 
states results in replacing $\epsilon^{i}(\lambda_{g})$ with 
$\epsilon^{i}_{c}({\bf Q},\lambda_{g})$,
where
\begin{equation}
 \epsilon^{i}_{c}({\bf Q},\lambda_{g})=\epsilon^{j}(\lambda_{g})
 (\delta^{ij}-\bar{Q}^{i}\bar{Q}^{j}).
 \end{equation}

Using the construction of spin wave functions summarized in Appendix B, 
the corresponding normalized wave functions are then given by 
\begin{equation} \Psi^{\lambda_{ex}}_{q\bar{q}g(S)}=
 \sum_{\lambda_{q\bar{q}}}\Psi^{\lambda_{q\bar{q}}}_{q\bar{q}}({\bf q},
 -{\bf Q},\lambda_{q},\lambda_{\bar{q}})\zeta_{(S)}(\bar{{\bf Q}},
 \lambda_{q\bar{q}},\lambda_{g},\lambda_{ex}), 
\label{wavef}
\end{equation}
where the spin states $\zeta_{(S)}$ in Eq.\ \ref{wavef} are given by 
\begin{equation}
 \zeta_{(S=0)}= \sqrt{\frac{3}{8\pi}}[{\bf \epsilon}^{\ast}
 (\lambda_{q\bar{q}})\cdot{\bf \epsilon}_{c}^{\ast}({\bf Q},\lambda_{g})]
 [\bar{{\bf Q}}\cdot{\bf \epsilon}(\lambda_{ex})], 
 \label{eq:wavef1} 
\end{equation}
\begin{equation} 
 \zeta_{(S=1)}=\sqrt{\frac{3}{8\pi}}[{\bf \epsilon}^{\ast}
 (\lambda_{q\bar{q}})\times{\bf \epsilon}_{c}^{\ast}({\bf Q},\lambda_{g})]
 \cdot[\bar{{\bf Q}}\times{\bf \epsilon}(\lambda_{ex})], 
\label{eq:wavef2} 
\end{equation}
\begin{equation} 
\zeta_{(S=2)}=\sqrt{\frac{27}{104\pi}}\bar{{\bf Q}}
 \cdot[{\bf \epsilon}^{\ast}(\lambda_{q\bar{q}})
 \otimes{\bf \epsilon}_{c}^{\ast}({\bf Q},\lambda_{g})]
 \cdot{\bf \epsilon}(\lambda_{ex}), 
\label{eq:wavef3}
\end{equation}
with $\bar{{\bf Q}}={\bf Q}/|{\bf Q}|$ and 
$(A\otimes B)_{ij}=2A_{(i}B_{j)}-\frac{2}{3}\delta_{ij}({\bf A}\cdot{\bf B})$.
It is easy to show 
$\zeta_{(2)}=3/\sqrt{13}(\zeta_{(1)}-\frac{2}{3}\zeta_{(0)})$. 
The loss of linear independence is directly related to the transversity of 
$\epsilon^{i}_{c}$.

The hybrid state in its rest frame is given by
\begin{widetext}
\begin{eqnarray}
|\pi_{1}(I_{3},\lambda_{ex})\rangle=& & \sum_{all\,\,\lambda,c,f}
 \frac{1}{N_{ex}}\int\frac{d^{3}{\bf p}_{q}\,d^{3}{\bf p}_{\bar{q}}
 \,d^{3}{\bf Q}}{(2\pi)^{9}2E_{q}2E_{\bar{q}}E_{g}} 
 \cdot(2\pi)^{3}2(E_{q}+E_{\bar{q}}+E_{g})\delta^{3}
 ({\bf p}_{q}+{\bf p}_{\bar{q}}+{\bf Q})
 \frac{\lambda^{c_{g}}_{c_{q}c_{\bar{q}}}}{2}
 \frac{\sigma^{i}_{f_{q}f_{\bar{q}}}\epsilon^{i}(I_{3})}{\sqrt{2}}  
 \nonumber  \\
& & \times \Psi^{\lambda_{ex}}_{q\bar{q}g}({\bf p}_{q},{\bf p}_{\bar{q}},
 \lambda_{q},\lambda_{\bar{q}},\lambda_{g})\psi_{L}'\,b^{\dag}_{{\bf p}_{q}
 \lambda_{q}f_{q}c_{q}}d^{\dag}_{{\bf p}_{\bar{q}}
 \lambda_{\bar{q}}f_{\bar{q}}c_{\bar{q}}}a^{\dag}_{{\bf Q}
 \lambda_{g}c_g}|0\rangle,
\end{eqnarray}
\end{widetext} 
where the spin wave function $\Psi_{q\bar{q}g}$ was given in Eq.\ 
($\ref{wavef}$) for $S=0,1,2$, and the orbital wave function $\psi_{L}'$ 
depends only on $m_{q\bar{q}}$ and the invariant mass of the three-body 
system, $m_{q{\bar q}g}$.  Here $m_{g}$ denotes the effective mass of the 
gluon and $E_{g}=E(m_{g},{\bf Q})$, while  
$\lambda^{c_{g}}_{c_{q}c_{\bar{q}}}$ are the Gell-Mann matrices. 

The orbital angular momentum wave function for the $Q\bar{Q}g$ system should 
depend only on the invariant masses $m_{q\bar{q}}$, $m_{q\bar{q}g}$, and 
we will again introduce an exponential function
\begin{equation}
 \psi'_{L}(m_{q\bar{q}}/\mu_{ex},m_{q\bar{q}g}/\mu'_{ex})=
 e^{-m^{2}_{q\bar{q}}/8\mu^{2}_{ex}}\cdot 
 e^{-m^{2}_{q\bar{q}g}/8\mu'^{2}_{ex}}. 
 \end{equation}
In the non-relativistic limit only that part of the $\pi_{1}$ spin wave 
function corresponding to the $Q\bar{Q}$ pair is reduced to a Clebsch-Gordan 
coefficient, whereas the functions $\zeta$ given in Eqs.\ 
($\ref{eq:wavef1}-\ref{eq:wavef3}$) remain unchanged.

\begin{figure}
 \includegraphics[width=3.in]{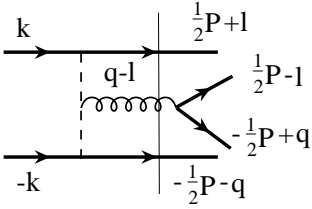}
 \caption{\label{fig3}
 Coulomb gauge description of normal meson decays. The $Q{\bar Q}$ component 
of the meson mixes with the $Q{\bar Q}g$ state with a subsequent decay of 
the transverse gluon into a $Q{\bar Q}$ pair. The vertical solid line 
represents the sum over all $Q{\bar Q}g$ intermediate states. }
\end{figure}

\subsection{Non-exotic hybrids or $Q\bar{Q}g$ components of normal mesons} 

As discussed in Section 1, in Coulomb gauge the decay of a normal 
$Q{\bar Q}$ meson is expected to proceed via its $Q{\bar Q}g$ component with 
the gluon dissociating into a $Q{\bar Q}$ pair, as shown in 
Fig.~\ref{fig3}.  The $Q{\bar Q}g$ component of the wave function is 
obtained by integrating the $Q{\bar Q}$ state over the amplitude of 
transverse gluon emission from the Coulomb line~\cite{ases2},  shown by the vertical dashed line in Fig.~\ref{fig3},  which gives,

\begin{widetext}
\begin{equation}
 \Psi_{q{\bar q}g}( {1\over 2}{\bf P} + {\bf l}, -{1\over 2}{\bf P} - {\bf q}, 
 {\bf q} - {\bf l} ) =   \int {{d^3{\bf k}}\over {(2\pi)^3}} \Psi_{q{\bar q}}
 ({\bf k}) V({\bf k} - {1\over 2} {\bf P} - {\bf l}, {\bf k} -{1\over 2}
 {\bf P} - {\bf q} )
{{ \epsilon_{c}({\bf q}-{\bf l},\lambda_g)\cdot  \left[ {\bf k} - {1\over 2}
 ({\bf P} + {\bf l} + {\bf q}) \right] } \over {
 \sqrt{2\omega_g({\bf q} - {\bf l})}     \Delta E} }  
  \label{qqg}
\end{equation}
\end{widetext}
In Eq.~(\ref{qqg}) $V(p,q)$ is given by a product of the two Faddeev-Popov 
operators (represented by the dashed line in Fig.\ \ref{fig3}) modified by a 
Coulomb kernel vertex correction. The transverse gluon couples to the 
Coulomb line via a derivative coupling which produces the momentum 
dependence of the numerator. The denominator is given by the difference 
between the energy of the $Q{\bar Q}$ state and the energy of the 
$Q{\bar Q}g$ hybrid (non-exotic) state in the absence of mixing between 
the two~\cite{aspk}.

The spin-orbit structure of the $Q{\bar Q}g$ component can be inferred from 
Eq.~(\ref{qqg}).  The momentum vector in the numerator can be coupled with 
the spin-orbit component of the $\Psi_{q\bar q}$ wave function and later 
coupled with the $J^{PC}=1^{--}$ transverse gluon.  For example for the 
$\rho$ meson the $Q{\bar Q}g$ component can be expanded in a basis of 
the $a_0$, $a_1$, $a_2$ - like $Q{\bar Q}$ wave functions all having spin one  
and one unit of orbital angular momentum between the quark and antiquark, 
coupled with the transverse gluon wave function 
$\epsilon_{c}({\bf q} - {\bf l},\lambda_g)$  to give  $J^{PC}=1^{--}$.  
Specifically for the $\rho$ meson one obtains, 
 \begin{eqnarray}
& & \Psi^{\lambda_{\rho}}_{q\bar{q}g(J)}(\lambda_{q},\lambda_{\bar{q}},
 \lambda_{g})=\sum_{\lambda_{q\bar{q}},\sigma}\Psi^{J,
 \lambda_{q\bar{q}}}_{q\bar{q}}({\bf q},{\bf l}_{q\bar{q}}=-{\bf Q},
 \lambda_{q},\lambda_{\bar{q}})  \nonumber \\
& & \times D^{(1)\ast}_{\lambda_{g}\sigma}(\bar{{\bf Q}}) \langle J,
 \lambda_{q\bar{q}};1,\sigma|1,\lambda_{\rho} \rangle .
\end{eqnarray}
Here $\Psi^{J,\lambda_{q\bar{q}}}_{q\bar{q}}$ are the $a_0, a_1$ and 
$a_2$ $Q{\bar Q}$ wave functions for $J=0,1,2$ respectively, and the gluon 
helicity $\sigma=\pm1$.  The normalized wave functions for $\rho$, similar 
to those of $\pi_{1}$ in Eq.\ ($\ref{wavef}$), are then given by
\begin{equation}
 \Psi^{\lambda_{\rho}}_{q\bar{q}g(J)}=\sum_{\lambda}
 \Psi^{\lambda}_{q\bar{q}}({\bf q},-{\bf Q},\lambda_{q},\lambda_{\bar{q}})
 \zeta_{(J)}(\bar{{\bf Q}},\lambda,\lambda_{g},\lambda_{\rho}), 
 \end{equation}
where 
\begin{eqnarray}
 \zeta_{(J=0)}&=&\sqrt{\frac{3}{8\pi}}[{\bf \epsilon}^{\ast}(\lambda)
 \cdot\bar{{\bf q}}][{\bf \epsilon}_{c}^{\ast}({\bf Q},\lambda_{g})
 \cdot{\bf \epsilon}(\lambda_{\rho})], \\ 
 \zeta_{(J=1)}&=&\sqrt{\frac{9}{32\pi}}[{\bf \epsilon}^{\ast}(\lambda)
 \times\bar{{\bf q}}]\cdot[{\bf \epsilon}_{c}^{\ast}({\bf Q},\lambda_{g})
 \times{\bf \epsilon}(\lambda_{\rho})], \\ 
\zeta_{(J=2)}&=&\sqrt{\frac{27}{160\pi}}{\bf \epsilon}_{c}^{\ast}({\bf Q},
 \lambda_{g})\cdot[{\bf \epsilon}^{\ast}(\lambda)\otimes\bar{{\bf q}}]
 \cdot{\bf \epsilon}(\lambda_{\rho}).
\label{eq:nz}
\end{eqnarray}
Here ${\bf q}$ denotes again the quark momentum in the rest frame of the 
$Q\bar{Q}$ pair.  The most general wave function will be given by a linear 
combination of the three components listed above, and it can be calculated 
from Eq.~(\ref{qqg}). 

We will also study decays of the $b_1$ meson which are often used as a 
testing ground for decay models. The $Q{\bar Q}g$ wave function with 
$J^{PC}=1^{+-}$, $I=1$ quantum numbers requires the $Q{\bar Q}$ to have 
$\pi$ or $\pi_2$ quantum numbers. The corresponding total wave functions 
are given by 
\begin{eqnarray}
 & &\Psi^{\lambda_{b_{1}}}_{q\bar{q}g(J)}(\lambda_{q},\lambda_{\bar{q}},
 \lambda_{g})= \sum_{\lambda_{q\bar{q}},\sigma}\Psi^{J,
 \lambda_{q\bar{q}}}_{q\bar{q}}({\bf q},{\bf l}_{q\bar{q}}=-{\bf Q},
 \lambda_{q},\lambda_{\bar{q}})  \nonumber \\
& & \times D^{(1)\ast}_{\lambda_{g}\sigma}(\bar{{\bf Q}}) \langle J,
 \lambda_{q\bar{q}};1,\sigma|1,\lambda_{b_1} \rangle,
\label{eq:b1}
\end{eqnarray}
with $\Psi^{J,\lambda_{q\bar{q}}}_{q\bar{q}}$ being the $\pi$ ($\pi_2$) 
$Q{\bar Q}$ wave function for $J=0$ ($J=2$).  
Thus the normalized spin wave functions are given by
\begin{equation}
 \Psi^{\lambda_{b_1}}_{q\bar{q}g(J)}=\sum_{\lambda}\Psi^{\lambda}_{q\bar{q}}
 ({\bf q},-{\bf Q},\lambda_{q},\lambda_{\bar{q}})\zeta_{(J)}(\bar{{\bf Q}},
 \lambda,\lambda_{g},\lambda_{\rho}),
 \label{34} 
 \end{equation}
 where 
 \begin{equation} 
\zeta_{(J=0)} = \sqrt{ 3\over {8\pi}} [\epsilon_{c}^{\ast}({\bf Q},
 \lambda_{g})\cdot\epsilon(\lambda_{b_{1}})]
\end{equation}
and 
 \begin{equation} 
\zeta_{(J=2)} = \sqrt{27\over {64\pi}} \bar{{\bf q}}\cdot[ 
 \epsilon_{c}^{\ast}({\bf Q},\lambda_{g})\otimes  \epsilon(\lambda_{b_{1}})]
 \cdot\bar{{\bf q}}.
\end{equation}

\section{Exotic meson decays}

\subsection{Relativistic $^{3}S_{1}$ model}

We assume that the transverse gluon in the $\pi_{1}$ creates a quark-antiquark 
pair and the hybrid decays into two mesons with momenta ${\bf P}$ and 
$-{\bf P}$.  Since the quark pair is emitted in the $S=1$, $L=0$ state 
this decay mechanism is also referred to as the $^{3}S_{1}$ model.  
The Hamiltonian $H$ of this process in Coulomb gauge is given by 
\begin{equation} 
H_{qqg} = \sum_{c,f}\int d^{3}{\bf x}\,\bar{\psi}_{c_{1}f}({\bf x})
 (g{\bf \gamma}\cdot{\bf A}^{c_{g}}({\bf x}))\psi_{c_{2}f}({\bf x})
 \frac{\lambda^{c_{g}}_{c_{1}c_{2}}}{2}.
\label{eq:H}
\end{equation}
In the constituent basis used here the single-particle quark and antiquark 
orbitals correspond to states of massive particles with relativistic 
dispersion relations, in which a running quark mass is approximated by a 
constant constituent mass, 
\begin{eqnarray}
& & \psi_{cf}({\bf x})= \sum_{\lambda}\int\frac{d^{3}{\bf k}}
 {(2\pi)^{3}2E(m,{\bf k})}[u({\bf k},\lambda)b_{{\bf k}\lambda cf}+ 
 \nonumber \\
& & +v(-{\bf k},\lambda)d^{\dag}_{-{\bf k}\lambda cf}]e^{i{\bf k}
 \cdot{\bf x}} .
 \end{eqnarray}
Similarly the gluon field ${\bf A}^{c_{g}}$ is expanded in a basis of 
transverse quasi-gluons, with a single particle wave function characterizing 
a state of mass $m_g \sim 600\mbox{ MeV}$, 
\begin{equation}
 \sum_{\lambda}\int\frac{d^{3}{\bf k}}{(2\pi)^{3}2E(m_{g},{\bf k})}
 [{\bf \epsilon}_{c}({\bf k},\lambda)a^{c_{g}}_{{\bf k}\lambda}+
 {\bf \epsilon}_{c}^{\ast}(-{\bf k},\lambda)a^{\dag c_{g}}_{-{\bf k}\lambda}]
 e^{i{\bf k}\cdot{\bf x}}. 
 \end{equation}
In Eq.~(\ref{eq:H}), $g$ is the strong coupling constant, later chosen to be 
of the order $10$, corresponding to $\alpha_s = O(1)$. 
The decay matrix element 
\begin{equation}
 \langle M_{1}({\bf P}), M_{2}(-{\bf P})|H|\pi_{1} \rangle =(2\pi)^{3}
 \delta^{3}({\bf P}-{\bf P})A({\bf P}) 
 \end{equation} 
(where $M$ denotes a meson) determines the decay amplitude $A$. 

For decays of $\pi_{1}$ into $\pi\eta$ or $\pi b_{1}$, the spin part of the 
amplitude $A$ is proportional to 
\begin{equation} 
B^{\mu j}\sum_{\lambda_{g}}\psi_{\mu(S)}({\bf Q},\lambda_{q},
 \lambda_{\bar{q}},\lambda_{g},\lambda_{ex})\epsilon^{j}_{c}({\bf Q},
 \lambda_{g}),
\label{eq:help1}
\end{equation}
where
\begin{equation}
\psi_{\mu(S)}=\sum_{\lambda_{q\bar{q}}}\zeta_{(S)}({\bf Q},
 \lambda_{q\bar{q}},\lambda_{g},\lambda_{ex})\epsilon_{\mu}(-{\bf Q},
 \lambda_{q\bar{q}}), 
\end{equation}
and
\begin{widetext}
\begin{equation}
 B^{\mu j}=Tr\Bigl[(\not{k}-m)(\not{p}-m)\Bigl(\gamma^{\mu}+
\frac{p^{\mu}-l^{\mu}}{m_{q\bar{q}}({\bf p},{\bf l})+2m}\Bigr) 
 (\not{l}+m)(\not{r}+m)\gamma^{j}\Bigr].
\label{eq:tr1}
\end{equation}
\end{widetext}
In this expression $p$ and $l$ are respectively the four-momenta of the quark 
and the antiquark in the $\pi_{1}$, whereas $r$ and $k$ are the four-momenta 
of the quark and the antiquark created from the gluon.
If $\mu_{\eta}=\mu_{\pi}$ then $A_{\pi\eta}=0$. The $1^{-+}$ state is also 
found to have vanishing decay amplitude into two pions because of a 
relative minus sign from isospin that makes both terms cancel.  
Thus we find that the $1^{-+}$ isovector does not decay to identical 
pseudoscalars.  This is a relativistic generalization of a 
symmetry found in other non-relativistic decay models~\cite{closeandpage}.  
Since $\pi$ and $\eta$ are to a good approximation members of the same 
flavor multiplet, in the quark model one typically finds their orbital wave 
functions to be similar, \IE  $\mu_{\pi}  \sim \mu_{\eta}$, resulting 
in a small $\pi_1 \to \eta \pi$ decay rate.  Of the two decay channels 
 $\pi\eta$ and $\pi b_{1}$, the latter will be favored.  However, 
the parameters $\mu$ need not be similar for two mesons with 
different radial quantum numbers, making corresponding channels significant. 
For decays of $\pi_{1}$ into $\pi\rho$, $\pi f_{J}$ or 
$\eta a_{J}$ ($J=0,2$), the spin part is proportional to
\begin{equation} 
C^{\mu\nu j}\sum_{\lambda_{g}}\psi_{\mu(S)}({\bf Q},\lambda_{q},
 \lambda_{\bar{q}},\lambda_{g},\lambda_{ex})\epsilon^{j}_{c}({\bf Q},
 \lambda_{g})\epsilon^{\nu\ast}(-{\bf P},\lambda),
\label{eq:help2}
\end{equation}
where $\lambda$ is the spin of the second meson, and 
\begin{widetext}
\begin{equation}
 C^{\mu\nu j}=\Bigl[(\not{p}+m)\Bigl(\gamma^{\mu}-\frac{p^{\mu}-l^{\mu}}
 {m_{q\bar{q}}({\bf p},{\bf l})+2m}\Bigr)(\not{l}-m) 
\Bigl(\gamma^{\nu}-\frac{r^{\nu}-l^{\nu}}{m_{q\bar{q}}({\bf r},
 {\bf l})+2m}\Bigr)(\not{r}+m)\gamma^{j}(\not{k}-m)\gamma^{5}\Bigr] \,.
\label{eq:tr2}
\end{equation}
\end{widetext} 
If $\mu_{\rho}=\mu_{\pi}$ the amplitude of the decay into $\pi\rho$ does 
not vanish (unlike $\pi\eta$) and this channel can be favored.  The same 
holds for $\pi_{1}\rightarrow\pi f_{J}$ and $\pi_{1}\rightarrow\eta a_{J}$.  
For the decay $\pi_{1}\rightarrow\rho\omega$, the spin part is proportional 
to
\begin{equation}
 D^{\mu\nu\rho j}\sum_{\lambda_{g}}\psi_{\mu(S)}({\bf Q},\lambda_{g})
 \epsilon^{j}_{c}({\bf Q},\lambda_{g})\epsilon^{\nu\ast}({\bf P},
 \lambda_{\rho})\epsilon^{\rho\ast}(-{\bf P},\lambda_{\omega}),
\label{eq:help3}
\end{equation}
where
\begin{widetext}
\begin{equation}
D^{\mu\nu\rho j}=Tr\Bigl[(\not{k}-m)\Bigl(\gamma^{\nu}-
 \frac{p^{\nu}-k^{\nu}}{m_{q\bar{q}}({\bf p},{\bf k})+2m}\Bigr)(\not{p}+m)
\Bigl(\gamma^{\mu}-\frac{p^{\mu}-l^{\mu}}{m_{q\bar{q}}({\bf p},{\bf l})+2m}
 \Bigr)(\not{l}-m)\Bigl(\gamma^{\rho}-\frac{r^{\rho}-l^{\rho}}{m_{q\bar{q}}
 ({\bf r},{\bf l})+2m}\Bigr)(\not{r}+m)\gamma^{j}\Bigr].
\label{eq:tr3}
\end{equation}
\end{widetext}
If $\mu_{\rho}=\mu_{\omega}$ then $A=0$ and the hybrid will not decay 
into $\rho$ and $\omega$.  Because both parameters $\mu$ are expected to be 
of the same order, the channel $\pi_{1}\rightarrow\rho\omega$ will not be 
favored. 

Finally, for decays into strange mesons, one should use the above spin 
factors (depending on the quantum numbers), with a small modification 
resulting from $r^{2}=k^{2}=m_{s}^{2}$.  The amplitudes $A_{KK_{1}}$ 
($S_{q\bar{q}}=0,1$) behave similarly to $A_{\pi b_{1}}$ and 
$A_{\pi f_{1}}$, whereas $A_{KK^{\ast}}$ is like $A_{\pi\rho}$. 
Therefore the former will be dominant and the latter is expected to be 
much smaller.

The width rate for a decay into a final state with orbital angular momentum $L$ is equal to
\begin{equation} 
\Gamma_{L}=\frac{P_{0}}{32\pi^{2}m^{2}_{ex}}a^{2}_{L}(P_{0}), 
\end{equation}
where $P_{0}$ is defined by 
$E(m_{1},{\bf P}_{0})+E(m_{2},-{\bf P}_{0})=m_{ex}$. 
The partial wave amplitudes are given by
\begin{widetext}
\begin{equation}
a_{L}(P)=\sum_{J,\lambda,M}\int A({\bf P},\lambda_{1},\lambda_{2},
\lambda_{ex}) \langle J_{1},\lambda_{1};J_{2},\lambda_{2}|J,\lambda \rangle 
 Y_{LM}({\bf P}) \langle J,\lambda;L,M|1,\lambda_{ex} \rangle d\Omega, 
 \end{equation}
\end{widetext}
with $d\Omega$ being the element of the solid angle in the direction of 
${\bf P}$, while $m$ and $\lambda$ are respectively the masses and spins of 
the outgoing mesons.

\subsection{Non-relativistic limit}

The non-relativistic limit is obtained when the quark masses are large 
compared to the quantities $\mu$ and $P_{0}$.  This is equivalent to 
ignoring the Wigner rotation and taking non-relativistic phase space.  
In the orbital wave functions, however, we must keep next to leading 
order terms that depend on momenta, otherwise the amplitude would become 
divergent.  For $\pi_{1}\rightarrow\pi\eta,\,\pi b_{1}$ the dominant 
term has the form 
\begin{equation}
 B^{ij}\rightarrow-32m^{4}\delta^{ij} 
 \end{equation}
while the other components are much smaller.  Therefore the spin factor 
of Eq.~($\ref{eq:help1}$) vanishes for $S=1$, and for $S=0$ it tends to 
$8\sqrt{24/\pi}m^{4}\bar{Q}^{l}\epsilon^{l}(\lambda_{ex})$.
From the above it follows $\Gamma_{(S=1)}\rightarrow0$ and 
$\Gamma_{(S=2)}/\Gamma_{(S=0)}\rightarrow4/13$. For 
$\pi_{1}\rightarrow\pi\rho,\,\pi f_{1}$ we have
\begin{equation}
 C^{ikj}\rightarrow-32im^{4}\epsilon^{0ikj} 
 \end{equation}
and the other components are much smaller.  Therefore the spin factor 
of Eq.~($\ref{eq:help2}$) vanishes for $S=0$, whereas for $S=1$ it tends to 
$-8\sqrt{6/\pi}im^{4}\bar{Q}^{i}\epsilon^{j}(\lambda_{ex})\epsilon^{k\ast}
(\lambda)\epsilon^{ikj}$. Thus $\Gamma_{(S=0)}\rightarrow0$ and 
$\Gamma_{(S=2)}/\Gamma_{(S=1)}\rightarrow9/13$.
Finally, for $\pi_{1}\rightarrow\rho\omega$ we have
\begin{equation}
 D^{ijkl}\rightarrow32m^{4}(\delta^{ij}\delta^{kl}-\delta^{ik}\delta^{jl}+
 \delta^{il}\delta^{jk}) 
 \end{equation}
(the other components are again much smaller). Therefore the spin factor 
of Eq.~($\ref{eq:help3}$) tends for $S=0$ to 
$8\sqrt{24/\pi}m^{4}\bar{Q}^{i}\epsilon^{i}(\lambda_{ex})\epsilon^{j\ast}
(\lambda_{\rho})\epsilon^{j\ast}(\lambda_{\omega})$, and for $S=1$ to 
$8\sqrt{6/\pi}m^{4}(\bar{Q}^{i}\delta^{jk}-\bar{Q}^{j}\delta^{ik})
\epsilon^{i\ast}(\lambda_{\rho})\epsilon^{j\ast}(\lambda_{\omega})
\epsilon^{k}(\lambda_{ex})$. 

We can straightforwardly understand the difference in amplitudes coming 
from the spin wave function.  If we assume $m_{\eta}=m_{\rho},\,\mu_{\eta}=
\mu_{\rho}$ and $\,m_{b_{1}}=m_{f_{1}}=m_{f_{2}},\,\mu_{b_{1}}=\mu_{f_{1}}
=\mu_{f_{2}}$ (the second condition for masses is satisfied to a good 
approximation), then one obtains 
\begin{equation}
A_{\pi\rho}=\frac{1}{2}A_{\pi\eta}\,\rightarrow\,\Gamma_{\pi\rho}
 =\frac{1}{2}\Gamma_{\pi\eta} 
\end{equation}
and
\begin{equation} 
A_{\pi f_{1}}=\frac{1}{2\sqrt{2}}A_{\pi b_{1}}\,\rightarrow\,
 \Gamma_{\pi f_{1}}=\frac{1}{8}\Gamma_{\pi b_{1}}, 
\end{equation}
where $A_{\pi\eta}$, $A_{\pi b_{1}}$ are taken for $S=0$ and 
$A_{\pi\rho}$, $A_{\pi f_{1,2}}$ for $S=1$. 
The relation between $A_{\pi\eta}$ and $A_{\pi b_{1}}$ (or between $A_{\pi\rho}$ and $A_{\pi f_{1,2}}$) is more complicated and 
depends on the orbital angular momentum wave functions 
$\psi_{L}$ and $\psi'_{L}$. 
If $\mu_{\rho}=\mu_{\pi}$ then in the non-relativistic limit $\pi_{1}$ 
will not decay into $\pi\rho$.  Therefore the width for this process 
is expected to be much smaller than that of $\pi b_{1}$, assuming the 
parameters $\mu_{\rho}$ and $\mu_{\pi}$ are very close to one another. 
Analogous calculations can be made for the decays of $\pi_{1}$ into strange 
mesons.  If $\mu_{K^{\ast}}=\mu_{K}$ then in the non-relativistic limit 
$A_{KK^{\ast}}=0$

\subsection{Numerical results}

Our model contains the following free parameters: the quark masses $m$;  
the effective gluon mass  $m_{g}$; the size parameters $\mu$ for 
the wave functions; and the strong coupling constant $g$.  The wave 
function parameters are constrained by the pion decay constant $f_{\pi}$ 
and the elastic form factor $F_{\pi}$. They are respectively given by 
\begin{equation}
 \langle 0|A^{\mu,i}({\bf 0})|\pi^{k}({\bf p}) \rangle =
 f_{\pi}p^{\mu}\delta_{ik} \, , 
 \end{equation}
and
\begin{equation}
 \langle \pi^{i}({\bf p}')|V^{\mu,j}({\bf 0})|\pi^{k}({\bf p}) \rangle =
 F_{\pi}(p^{\mu}+p'^{\mu})i\epsilon_{ijk} \,.
\label{eq:formf1} 
\end{equation}
The axial current $A^{\mu,i}({\bf 0})$ is equal to 
$\bar{\psi}_{cf}({\bf 0})\gamma^{\mu}\gamma_{5}\sigma^{i}\psi_{cf}/2$ and 
the vector current $V^{\mu,j}({\bf 0})$ to $\bar{\psi}_{cf}({\bf 0})
\gamma^{\mu}\sigma^{j}\psi_{cf}/2$. 
By virtue of the Lorentz invariance $f_{\pi}$ is a constant, whereas 
$F_{\pi}$ is a function of $Q^{2}=-({\bf p}-{\bf p}')^{2}$.  
As was discussed previously, it is not possible to construct the currents 
and wave functions with a fixed number of constituents in a Lorentz covariant 
manner. Thus the current matrix elements are expected to violate Lorentz 
covariance.  This will be reflected, for example in different values of 
$f_\pi$ obtained from spatial and time components of the axial current  
(rotational symmetry is not broken).  Even if we replaced the factor 
$E(m_{M},{\bf P})$ in Eq.\ ($\ref{eq:norm}$) by $1$, it would be very 
difficult to find generators of the Poincare group that satisfy the 
exact commutation relations. Thus our model with exponential orbital 
wave functions will not be completely invariant.  The resulting form factors 
will depend on the frame of reference and in order to obtain  
$F_{\pi}(Q^{2}=0)=1$ one typically employs the time component $\mu=0$ and 
works in the Breit frame.  For other light unflavored mesons we do not 
expect much variation on the ground state wave function and we choose the 
scale parameter $\mu$ to be of the same order as $\mu_{\pi}$. For kaons the 
same procedure fits $m_{s}$ and $\mu_{K}$. 

\begin{figure}
\includegraphics[width=2.0in]{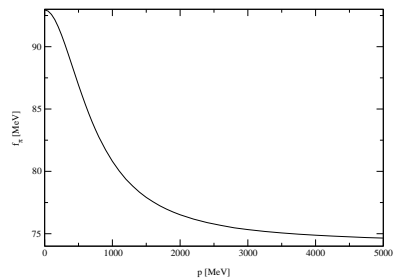}
\caption{\label{fig4} The pion weak decay constant $f_{\pi}$ as a function of 
the pion momentum $p$ for $m=306\mbox{ MeV}$ and $\mu_{\pi}=221\mbox{ MeV}$.}
\end{figure}

We will assume that the mass difference between $\pi$ and $\rho$ arises only 
from spin.  Therefore we can write
\begin{equation}
 m_{M}={\bar{m}}_{M}+k(s_{1}\cdot s_{2}), 
 \end{equation}
where $M$ denotes either meson and ${\bar{m}}_{M}$ is its 'averaged' mass. 
Substituting $m_{\pi}=140\mbox{ MeV}$ and $m_{\rho}=770\mbox{ MeV}$ we 
obtain ${\bar{m}}_{M}=612\mbox{ MeV}$, and thus for the constituent quark 
masses  we choose  $m_{u}=m_{d}={\bar{m}}_{M}/2=306\mbox{ MeV}$. 
A similar relation can be used for the $K$ and $K^{\ast}$ mesons, leading to 
${\bar{m}}_{K}=792\mbox{ MeV}$ and 
$m_{s}={\bar{m}}_{K}-m_{u}=486\mbox{ MeV}$.  The 'averaged' masses should 
be used in the normalization constants. 

\begin{figure}
\includegraphics[width=1.5in,angle=0]{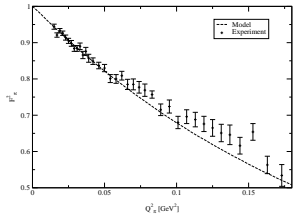}
\caption{\label{fig5} The pion electromagnetic form factor $F^{2}_{\pi}$ as a 
function of the momentum transfer $Q^{2}$ for $m=306\mbox{ MeV}$ and 
$\mu_{\pi}=221\mbox{ MeV}$. The experimental values are from 
Ref.~\cite{pipi-data}. }
\end{figure}

The weak decay constants can be used to fit the parameters $\mu_{\pi}$ and 
$\mu_{K}$.  Because of the Lorentz covariance breaking mentioned before, 
they become a function of the meson momentum and we choose them to be 
equal to their experimental values at rest.
Thus setting $f_{\pi}(0)=93\mbox{ MeV}$ and $f_{K}(0)=113\mbox{ MeV}$, 
leads to $\mu_{\pi}=221\mbox{ MeV}$ and $\mu_{K}=275\mbox{ MeV}$.  The 
momentum dependence of $f_{\pi}$ in our model is presented in 
Fig.~\ref{fig4}, which shows $\sim 20\%$ difference between the value of $f_{\pi}$ calculated for the meson at rest and for the meson with momentum approaching the light cone. 
In Fig.~\ref{fig5} we present $F^{2}(Q^{2})$ calculated with the wave 
function parameters obtained above and compared with data.  There is good 
agreement for small momentum transfer; the discrepancy for larger $Q^{2}$ 
indicates the lack of sufficient high momentum components in our wave 
function.  The strong coupling constant at this scale is approximately 
$g^{2}=10$, and we take $m_{g}=500\mbox{ MeV}$ for the gluon effective mass.  
Around this value the Coulomb interaction between quark and antiquark appears 
to be linear.

Next we proceed to discuss relativistic effects in $\pi_1$ exotic meson 
decays.  In Tables I and II we present the rates for various decay channels.  
The numbers in parentheses correspond to calculations using the 
non-relativistic formulae, where $S$ denotes the total spin of the 
$Q{\bar Q}g$ component of the $\pi_{1}$ wave function.  For all 
unflavored mesons and the $\pi_1$, the value of the parameter $\mu$ was 
taken equal to $\mu_{\pi}$, and for all strange mesons $\mu$ was set 
equal to $\mu_{K}$.  This assumption makes the widths for the channels 
$\pi\eta$, $\pi\eta'$ and $\rho\omega$ identically equal to zero. 

\begin{table}
\centering
\begin{tabular}{|c|c||r|r|r|}
\hline
\multicolumn{2}{|c||}{$\Gamma_{rel}(\Gamma_{nrel})$}&$S=0$&$S=1$&$S=2$\\
\hline\hline
$\pi b_{1}(1235)$&S& $150(259)$ & $<1(0)$ & $44(80)$\\
\cline{2-5}
 &D& $<1(<1)$ & $<1(0)$ & $<1(<1)$\\
\hline
$\pi f_{1}(1285)$&S& $<1(0)$ & $20(33)$ & $14(23)$\\
\cline{2-5}
 &D& $<1(0)$ & $<1(<1)$ & $<1(<1)$\\
\hline
$\pi f_{2}(1270)$&D& $<1(0)$ & $<1(<1)$ & $<1(<1)$\\
\hline
$\pi\rho(770)$&P& $3(0)$ & $<1(0)$ & $1(0)$\\
\hline
$KK^{\ast}(892)$&P& $1(0)$ & $<1(0)$ & $<1(0)$\\
\hline
\end{tabular}
\caption{\label{tab1} Relativistic (non-relativistic) widths in MeV for 
decays of the $\pi_{1}(1600)$. }
\end{table}

\begin{table}
\centering
\begin{tabular}{|c|c||r|r|r|}
\hline
\multicolumn{2}{|c||}{$\Gamma_{rel}(\Gamma_{nrel})$}&$S=0$&$S=1$&$S=2$\\
\hline\hline
$\pi b_{1}(1235)$&S& $48(70)$ & $<1(0)$ & $13(22)$\\
\cline{2-5}
 &D& $1(2)$ & $<1(0)$ & $<1(<1)$\\
\hline
$\pi f_{1}(1285)$&S& $<1(0)$ & $7(11)$ & $5(8)$\\
\cline{2-5}
 &D& $<1(0)$ & $2(<1)$ & $1(<1)$\\
\hline
$\pi f_{2}(1270)$&D& $<1(0)$ & $2(<1)$ & $1(<1)$\\
\hline
$\pi\rho$&P& $2(0)$ & $<1(0)$ & $<1(0)$\\
\hline
$\eta a_{1}(1260)$&S& $<1(0)$ & $13(22)$ & $9(16)$\\
\cline{2-5}
 &D& $<1(0)$ & $1(<1)$ & $<1(<1)$\\
\hline
$\eta a_{2}(1320)$&D& $<1(0)$ & $1(<1)$ & $<1(<1)$\\
\hline
$KK_{1}(1400)$&S& $127(45)$ & $<1(0)$ & $39(14)$\\
\cline{2-5}
 &D& $<1(<1)$ & $<1(0)$ & $<1(<1)$\\
\hline
$KK_{1}(1270)$&S& $<1(0)$ & $11(4)$ & $7(3)$\\
\cline{2-5}
 &D& $<1(0)$ & $<1(<1)$ & $<1(<1)$\\
\hline
$KK^{\ast}(892)$&P& $1(0)$ & $<1(0)$ & $<1(0)$\\
\hline
\end{tabular}
\caption{\label{tab2} Relativistic (non-relativistic) widths in MeV for 
decays of the $\pi_{1}(2000)$. }
\end{table}

\begin{figure}
\includegraphics[width=2.0in,angle=0]{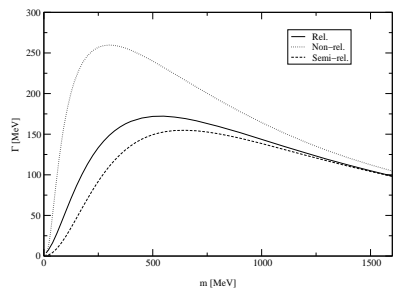}
\caption{\label{fig6} Relativistic, non-relativistic and semi-relativistic 
widths for $\pi_{1}\rightarrow\pi b_{1}$ in the S-wave state as a function 
of $m$, for $S=0$ and $m_{ex}=1600\mbox{ MeV}$. }
\end{figure}

\begin{figure}
\includegraphics[width=2.0in,angle=0]{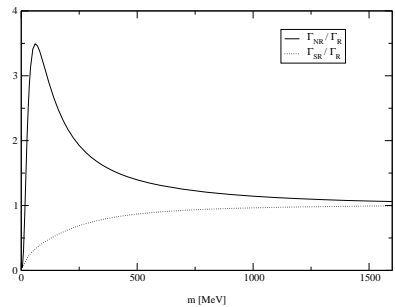}
\caption{\label{fig7} Ratios of non-relativistic to relativistic, and 
semi-relativistic to relativistic width rates for 
$\pi_{1}\rightarrow\pi b_{1}$ in the S-wave as a function of $m$, for 
$S=0$ and $m_{ex}=1600\mbox{ MeV}$. }
\end{figure}

In Fig.~\ref{fig6} we compare relativistic and non-relativistic 
predictions for the width for the decay $\pi_{1}\rightarrow\pi b_{1}$ as a 
function of the mass of the light quark $m$. It also shows the 
semi-relativistic values, which include relativistic phase space and 
orbital wave functions, but no Wigner rotations. The ratios of 
non-relativistic to relativistic (and semi-relativistic to relativistic) 
values are shown in Fig.~\ref{fig7}. 

From these results it is clear that fully relativistic results are 
significantly different from non-relativistic ones. There are two sources 
of this difference: the Wigner rotation which introduces relativistic 
coupling between spin and spatial degrees of freedom in the wave functions, 
and different relations between energy, momentum and the invariant masses (in 
the phase space and orbital wave functions). For realistic quark masses, 
both corrections appear to introduce corrections as large as 10\% 
and thus should be included in phenomenological models.

\section{Normal meson decays}

In this section we will calculate the widths of the decays 
$\rho\rightarrow2\pi$ and $b_{1}\rightarrow\pi\omega$.  These are the 
dominant decay channels (accounting for almost $100\%$ of the total width) 
and their values are well known from experiment, and therefore can be used 
to test the model presented in this work. 

As discussed previously, in Coulomb gauge the decay of a normal meson 
is expected to proceed via $Q{\bar Q}$ mixing with  the $Q{\bar Q}g$ hybrid 
component followed by gluon dissociation to a $Q{\bar Q}$ pair. However, the 
common approach to normal meson decays is based on the $^3P_0$ 
model where $Q{\bar Q}$ pair creation is described by an effective 
operator that creates the pair from the vacuum, in the presence of the normal 
$Q{\bar Q}$ component of the decaying meson. We will first discuss the role 
of relativistic effects in the $^3P_0$ model and then compare with  
predictions based on the Coulomb gauge picture.

\subsection{The decay $\rho\rightarrow\pi\pi$}

We start from the $^{3}P_{0}$ Hamiltonian
\begin{equation}
H=\Lambda\sum_{c,f}\int d^{3}{\bf x}\,\bar{\psi}_{cf}({\bf x})
 \psi_{cf}({\bf x}),
\label{eq:H2}
\end{equation}
where $\Lambda$ is a mass scale that can be fixed by the absolute decay 
width, and is expected to be of the order of the average quark momentum. 
In the decay matrix element the spin factor is proportional to 
\begin{eqnarray}
& & W^{\lambda_{\rho}}=Tr\Bigl[(\not{k}+m)(\not{p}+m)
 \Bigl(\gamma^{i}-\frac{p^{i}-l^{i}}{m_{q\bar{q}}({\bf p},{\bf l})+2m}\Bigr) 
  \nonumber \\
& & \times (\not{l}-m)(\not{r}-m)\Bigr]\epsilon^{i}(\lambda_{\rho}), 
\end{eqnarray}
with the same notation used in the previous section.  In the non-relativistic 
limit the above expression tends to 
$32m^{3}(p^{i}-P^{i})\epsilon^{i}(\lambda_{\rho})$.  The expression for 
this process is much simpler than for decays of $\pi_{1}$ because there is 
no gluon.

The decay of a $\rho$ treated as a gluonic bound state has a similar 
structure to the decay $\pi_{1}\rightarrow\pi\eta$, and proceeds via the 
$^{3}S_{1}$ interaction given by Eq.~(\ref{eq:H}).  The spin factor is the 
same as in Eq.~($\ref{eq:help1}$) with $S=J$ and $\lambda_{ex}$ replaced by 
$\lambda_{\rho}$, but the functions $\zeta$ are now given 
by Eqs.~(30-32).  In the non-relativistic limit this factor will become 
$-8\sqrt{2/\pi}m^{4}\bar{q}^{i}
\epsilon^{j}(\lambda_{\rho})(\delta^{ij}-\bar{Q}^{i}\bar{Q}^{j})$ for J=0, 
$-8\sqrt{9/2\pi}m^{4}\bar{q}^{i}\epsilon^{j}(\lambda_{\rho})
(\delta^{ij}+\bar{Q}^{i}\bar{Q}^{j})$ for J=1, and
$-8\sqrt{27/10\pi}m^{4}\bar{q}^{i}\epsilon^{j}(\lambda_{\rho})
\frac{1}{3}(7\delta^{ij}-\bar{Q}^{i}\bar{Q}^{j})$ for J=2.  None of these 
functions vanishes, but only two of them remain linearly independent. 

In Table~III, we present numerical predictions for the widths of this 
decay.  The experimental value of the width for $\rho\rightarrow2\pi$ is 
$149\mbox{ MeV}$. This number can be used to fit the free parameter 
$\Lambda$, the coupling constant $g$, or the hybrid scale parameter 
$\mu_{ex'}$.

\begin{table}
\vskip 20pt
\centering
\begin{tabular}{|c|c||r|r|r|r|}
\hline
\multicolumn{2}{|c||}{$\Gamma_{rel}(\Gamma_{nrel})$}&$^{3}P_{0}$
&$a_{0}$&$a_{1}$&$a_{2}$\\
\hline\hline
$2\pi$&P& $59(195)$ & $7(15)$ & $12(45)$ & $58(75)$\\
\hline
\end{tabular}
\caption{\label{tab3} Relativistic (non-relativistic) widths in MeV for 
the decay $\rho(770)\rightarrow\pi\pi$ for $\Lambda=\mu_{\pi}$. }
\end{table}

\subsection{The decay $b_{1}\rightarrow\pi\omega$}

This process is a better test for the different decay schemes, because the 
ratio of D-wave to S-wave widths is independent of the values of 
$\Lambda$ and $g$.  We will begin with the picture of the $b_{1}$ as a 
$Q\bar{Q}$ bound state and the decay Hamiltonian of Eq.~($\ref{eq:H2}$), 
\IE the $^3P_0$ model.  The spin factor is proportional to
\begin{eqnarray}
 & & W^{\lambda_{\omega}}=Tr\Bigl[(\not{r}+m)(\not{k}-m)(\not{p}-m)
(\not{l}-m) \nonumber \\
 & &  \times\Bigl(\gamma^{\mu}-\frac{r^{\mu}-l^{\mu}}{m_{q\bar{q}}({\bf r},
{\bf l})+2m}\Bigr)\Bigr]\epsilon^{\ast}_{\mu}(-{\bf P},\lambda_{\omega}), 
 \end{eqnarray}
with the same notation as for $\rho\rightarrow2\pi$.  In the non-relativistic 
limit this expression tends to 
$32m^{3}(p^{i}-P^{i})\epsilon^{i\ast}(\lambda_{\omega})$.
The spin factor for the decay of a $b_{1}$ treated as a gluonic bound state is given by Eq.~(\ref{eq:help1}) in which
\begin{widetext}
\begin{equation}
 B^{\mu j}=Tr\Bigl[(\not{k}-m)(\not{p}-m)(\not{l}-m)\Bigl(\gamma^{\mu}-
\frac{r^{\mu}-l^{\mu}}{m_{q\bar{q}}({\bf r},{\bf l})+2m}\Bigr)
 (\not{r}+m)\gamma^{j}\Bigr],
\label{eq:tr4}
\end{equation}
\end{widetext}
with $S=J$, $\lambda_{ex}$ replaced by $\lambda_{b_{1}}$, and the functions $\zeta$ given by Eqs.~(35-36).

\begin{table}
\centering
\begin{tabular}{|c|c||r|r|}
\hline
\multicolumn{2}{|c||}{$\Gamma_{rel}(\Gamma_{nrel})$}&$^{3}P_{0}$&$\pi$\\
\hline\hline
$\pi\omega(782)$&S& $16(15)$ & $52(82)$\\
\cline{2-4}
 &D& $17(42)$ & $<1(<1)$\\
\hline
\end{tabular}
\caption{\label{tab4} Relativistic (non-relativistic) widths in MeV of the 
decay $b_{1}(1235)\rightarrow\pi\omega(782)$ for $\Lambda=\mu_{\pi}$. }
\end{table}

The numerical predictions for the decay widths are presented in Table~IV.
The experimental value of the total width for the process 
$b_{1}\rightarrow\pi\omega$ is $142\mbox{ MeV}$, and the experimental ratio 
of the D-wave to S-wave widths is $0.08$.  Our predictions give a value 
less than $0.02$ for this ratio in the $^{3}S_{1}$ model, and close to $1$ 
for the $^{3}P_{0}$ decay.  The real mechanism for this decay appears to 
lie somewhere between the two predictions.  The $Q{\bar Q}$g wave function 
component of the $b_1$ wave functions used here is that of Eq.~(\ref{34}), 
corresponding to a $Q{\bar Q}$ pair with pion quantum numbers. 
For a $Q{\bar Q}$ pair with $\pi_2$ quantum numbers the numerical value for 
the width is much smaller than $1\mbox{ MeV}$ for the S-wave and 
approximately $1\mbox{ MeV}$ for the D-wave. The ratio $D/S$ is 
respectively $230$. In the non-relativistic limit one obtains similar results.
It is clear that treating $b_{1}$ as $\pi_{2}+g$ increases dramatically the 
$D/S$ ratio; therefore this may be an important component of the wave 
function. The relatively small values of the decay widths of a $b_{1}$ 
with $\pi_{2}$ quantum numbers (L=2) compared to those of 
a $b_{1}$ with pion quantum numbers (L=0) remind the situation for the 
process $\pi_{1}\rightarrow\pi b_{1}$, 
whose D-wave width was small compared to the S-wave.

\section{Summary and Outlook}

In this work we studied relativistic effects for the decays of normal and 
exotic mesons, and discussed a new picture of meson decays in the Coulomb 
gauge point of view. In Coulomb gauge the gluon degrees of freedom are 
physical.  Since they carry color, isolated gluons do not appear in the 
physical spectrum, and colorless excitations of $Q{\bar Q}g$ states are 
expected to be suppressed by a mass gap of the order of $1\mbox{ GeV}$. 
This is what lattice QCD studies find for the energy of gluonic excitations 
in the presence of $Q{\bar Q}$ sources.  Since strong decays are expected 
to proceed via gluon decay into a $Q{\bar Q}$ pair, the Coulomb gauge 
provides a natural framework for disentangling the dynamics of bound 
state formation (via a static Coulomb potential) and the decay of the 
gluonic component of a state. 

This work led to two important conclusions.  First, numerical results 
showed significant relativistic corrections arising from spin-orbit 
correlations introduced by Wigner rotation. The widths calculated using 
fully relativistic formulae are in general larger than the corresponding 
values calculated with no Wigner rotation (by a factor of roughly 
10\%), but smaller than completely non-relativistic rates.  Some decays that 
are suppressed in the non-relativistic limit (for example 
$\pi_{1}\rightarrow\pi\rho$ assuming identical orbital wave functions for 
$\pi$ and $\rho$) acquire non-zero amplitudes in the relativistic case.   

The second conclusion from this work is that the lightest exotic meson, the 
$\pi_{1}$, prefers to decay into two mesons, one of which has no orbital 
angular momentum while the other has $L=1$ (the so-called $S+P$ selection 
rule)~\cite{closeandpage}. 
Thus this selection rule, also found in other models seems to be quite 
robust~\cite{IKP}. 
Some decays ($\pi\eta$, $\rho\omega$, $KK^{\ast}$) are suppressed by 
symmetries in orbital wave functions, or the assumption that the parameter  
$\mu$ should be almost equal for mesons with the same radial quantum 
numbers.  We have also noticed that, for decays where two waves are 
possible, the rates for the higher partial waves are larger in the $^3P_0$ 
than in the $^3S_1$ model.
However, there is one caveat that needs to be 
explored further. There are several components to the $Q{\bar Q}g$ normal 
meson wave functions and if there are sizable contributions from wave 
functions with large relative angular momentum in the $Q{\bar Q}$ system, 
it is possible to obtain large amplitudes for high partial waves. 
This has in particular been illustrated in the case of the D-wave/S-wave 
ratio for the $b_{1}\rightarrow\pi\omega$ decay. 

The process $\pi_{1}\rightarrow\pi\rho$ seems to be suppressed and this agrees
with the $S+P$ selection rule~\cite{IKP}. However, some models predict larger 
values for its width~\cite{closeanddudek}. 
It is possible that these are increased by the final state interactions 
between the outgoing mesons.  The most important contribution may come 
from the process $\pi b_{1}\rightarrow\pi\rho$, that proceeds through 
an $\omega$ exchange.  Calculations of this contribution will be the subject 
of future work.

\begin{acknowledgments}
The authors wish to thank J.\ Narebski and S.\ Glazek for several 
discussions.  This work was supported in part by the US Department of 
Energy under contract DE-FG0287ER40365 and National Science Foundation 
grant NSF-PHY0302248.
\end{acknowledgments}

\section{Appendix}

\subsection{Boosted spin wave functions for mesons}

The polarization vectors corresponding to spin 1 quantized along the 
z-axis are given by
\begin{equation}
{\bf \epsilon}(\pm1)=\frac{\mp1}{\sqrt{2}}\left(\matrix{1\cr\pm i\cr0\cr}
 \right),\,\,{\bf \epsilon}(0)=\left(\matrix{0\cr0\cr1\cr}\right).
\label{eq:polar}
\end{equation}
The Wigner rotation matrix corresponding to a boost with 
${\bf \beta}\gamma={\bf P}/M$ is given by
\begin{widetext}
\begin{equation}
D^{(1/2)}_{\lambda\lambda'}({\bf q},{\bf P})=\Bigl[\frac{(E(m,{\bf q})+m)
 (E(M,{\bf P})+M)+{\bf P}\cdot{\bf q}+i{\bf \sigma}
 \cdot({\bf P}\times{\bf q})}{\sqrt{2(E(m,{\bf q})+m)(E(M,{\bf P})+M)
 (E(m,{\bf q})E(M,{\bf P})+{\bf P}\cdot{\bf q}+mM)}}\Bigr]_{\lambda\lambda'},
\end{equation}
\end{widetext}
whereas $S({\bf l}_{q\bar{q}}\rightarrow0)$ is the Dirac representation of 
the boost taking ${\bf l}_{q}$ to ${\bf q}$ and ${\bf l}_{\overline{q}}$ 
to $-{\bf q}$:
\begin{widetext}
\begin{equation}
S({\bf l}_{q\bar{q}}\rightarrow0)=\frac{1}{\sqrt{2m_{q\bar{q}}
 (E(m_{q\bar{q}},{\bf l}_{q\bar{q}})+m_{q\bar{q}})}}
 \left(\matrix{E(m_{q\bar{q}},{\bf l}_{q\bar{q}})+
 m_{q\bar{q}}&-{\bf \sigma}\cdot{\bf l}_{q\bar{q}}\cr -{\bf \sigma}
 \cdot{\bf l}_{q\bar{q}}&E(m_{q\bar{q}},{\bf l}_{q\bar{q}})+m_{q\bar{q}}\cr}
 \right).
\end{equation}
\end{widetext}
We make use of the relations:
\begin{equation}
\sum_{\sigma_{\bar{q}}}D^{(1/2)}_{\lambda_{\bar{q}}\sigma_{\bar{q}}}
 (-{\bf q},{\bf l}_{q\bar{q}})v(-{\bf q},\sigma_{\bar{q}})= 
 S({\bf l}_{q\bar{q}}\rightarrow0)v({\bf l}_{\bar{q}},\lambda_{\bar{q}}) \, , 
\end{equation}
\begin{equation}  
\sum_{\sigma_{q}}D^{\ast(1/2)}_{\lambda_{q}\sigma_{q}}({\bf q},
 {\bf l}_{q\bar{q}})u^{\dag}({\bf q},\sigma_{q})= u^{\dag}({\bf l}_{q},
 \lambda_{q})S^{\dag}({\bf l}_{q\bar{q}}\rightarrow0),
\end{equation}
and the formulae: 
\begin{eqnarray*}
 S^{\dag}\gamma^{0}&=&\gamma^{0}S^{-1}, \\ 
 S^{-1}\gamma^{i}S&=& \Lambda^{i}_{\,\,\nu}\gamma^{\nu} \\ 
 \epsilon_{\nu}({\bf l}_{q\bar{q}})&=&\Lambda_{\nu}^{\,\,i}
 (0\rightarrow{\bf l}_{q\bar{q}})\epsilon_{i}(\lambda_{q\bar{q}}) \, , 
\end{eqnarray*} 
one obtains Eq.~(\ref{eq:01}) from Eq.~(\ref{eq:01r}) and 
Eq.~(\ref{eq:boost}).  One can derive the spin wave functions 
for other mesons in similar fashion.

\subsection{Spin wave function of the $\pi_{1}$}

The transverse gluon states in the helicity basis $\sigma$ are related 
to the states in the spin basis $\lambda_{g}$ (quantized along a fixed 
z-axis) by
\begin{equation}
|{\bf Q},\lambda_{g}\rangle\,=\sum_{\sigma}D^{(1)\ast}_{\lambda_{g}\sigma}
 (\phi,\theta,-\phi)|{\bf Q},\sigma\rangle,
\end{equation}
where $\theta$ and $\phi$ are respectively the polar angle and azimuth of 
the gluon momentum direction.  For the gluon polarization vector we obtain
\begin{equation}
\epsilon^{i}_{c}({\bf Q},\lambda_{g})=\sum_{\sigma=\pm1}
 D^{(1)\ast}_{\lambda_{g}\sigma}(\phi,\theta,-\phi)\epsilon^{i}_{h}
 ({\bf Q},\sigma),
\end{equation}
where the helicity polarization vectors are given by
\begin{equation}
\epsilon^{i}_{h}({\bf Q},\sigma)=\sum_{\lambda_{g}}D^{(1)}_{\lambda_{g}\sigma}
 (\phi,\theta,-\phi)\epsilon^{i}(\lambda_{g}).
\end{equation}
Using the unitarity of the matrix $D^{(1)}$ one can show
\begin{equation}
\epsilon^{i}_{c}({\bf Q},\lambda_{g})\epsilon^{\ast i}_{h}({\bf Q},\sigma)
 =D^{(1)\ast}_{\lambda_{g}\sigma}(\bar{\bf{Q}}),
\end{equation}
and with the help of the identity 
\begin{eqnarray*} 
\epsilon^{\ast i}_{h}({\bf Q},\sigma)\epsilon^{j}_{h}({\bf Q},\sigma)
 &=&\delta^{ij}-\bar{Q}^{i}\bar{Q}^{j} 
\end{eqnarray*} 
we arrive at the result:
\begin{equation}
\epsilon^{i}_{c}({\bf Q},\lambda_{g})=\epsilon^{j}(\lambda_{g})
 (\delta^{ij}-\bar{Q}^{i}\bar{Q}^{j}),
\end{equation}
where $\bar{{\bf Q}}^{i}={\bf Q}^{i}/|{\bf Q}|$.

The Clebsch-Gordan coefficients and the spherical harmonic in 
Eq.~($\ref{eq:exo}$) can be expressed in terms of the polarization 
vectors ($\ref{eq:polar}$), for example:
\begin{eqnarray}
\langle 1,\lambda';0,0|1,\lambda \rangle &=&\epsilon^{\ast}(\lambda')
 \cdot\epsilon(\lambda), \\ 
 \langle 1,\lambda';1,\lambda|0,0 \rangle &=&\epsilon^{\ast}(\lambda')
 \cdot\epsilon^{\ast}(\lambda), \\ 
\langle 1,\lambda';1,\lambda''|1,\lambda \rangle &=& \frac{i}{\sqrt{2}}
 [\epsilon^{\ast}
 (\lambda')\times\epsilon^{\ast}(\lambda'')]\cdot\epsilon(\lambda), \\ 
Y_{1l}(\bar{{\bf Q}})&=&\sqrt{\frac{3}{4\pi}}\epsilon(l)\cdot\bar{{\bf Q}}.
\end{eqnarray}
Therefore we obtain:
\begin{widetext}
\begin{eqnarray}
\sum_{l} \langle 1,\lambda_{q\bar{q}};1,\lambda_{g}|0,0 
 \rangle Y_{1l}(\bar{{\bf Q}}) \langle 0,0;1,l|1,\lambda_{ex} 
 \rangle \,\,\,&=&[\epsilon^{\ast}(\lambda_{q\bar{q}})\cdot\epsilon^{\ast}
 (\lambda_{g})][\bar{{\bf Q}}\cdot\epsilon(\lambda_{ex})], \\ 
\sum_{l,s} \langle 1,\lambda_{q\bar{q}};1,\lambda_{g}|1,s \rangle Y_{1l}
 (\bar{{\bf Q}}) \langle 1,s;1,l|1,\lambda_{ex} \rangle \,\,\,&=&
 [\epsilon^{\ast}(\lambda_{q\bar{q}})\times\epsilon^{\ast}(\lambda_{g})]
 \cdot[\bar{{\bf Q}}\times\epsilon(\lambda_{ex})], \\ 
\sum_{l,s} \langle 1,\lambda_{q\bar{q}};1,\lambda_{g}|2,s \rangle 
 Y_{1l}(\bar{{\bf Q}}) \langle 2,s;1,l|1,\lambda_{ex} \rangle \,\,\,&=&
 \bar{{\bf Q}}\cdot[\epsilon^{\ast}(\lambda_{q\bar{q}})\otimes
 \epsilon^{\ast}(\lambda_{g})]\cdot\epsilon(\lambda_{ex}).
\end{eqnarray}
\end{widetext}
The action of the rotation matrix $D^{(1)}$ on the gluon states results in 
replacing $\epsilon^{i}(\lambda_{g})$ with 
$\epsilon^{i}_{c}({\bf Q},\lambda_{g})$ and that leads to the spin wave 
functions given in Eqs.~(\ref{eq:wavef1}-\ref{eq:wavef3}).

\end{document}